\documentclass{elsart3p}	

\usepackage{ifpdf}
\usepackage{color}

\usepackage{amssymb}
\usepackage{lineno}

\usepackage{graphicx}
\usepackage{subfigure}
\usepackage{captcont}

\journal{Astroparticle Physics}

\begin{document}

\begin{frontmatter}



\title{Measurement of acoustic attenuation in South Pole ice}

\bigskip



\author[Madison]{R.~Abbasi},
\author[Gent]{Y.~Abdou},
\author[RiverFalls]{T.~Abu-Zayyad},
\author[Christchurch]{J.~Adams},
\author[Madison]{J.~A.~Aguilar},
\author[Oxford]{M.~Ahlers},
\author[Madison]{K.~Andeen},
\author[Wuppertal]{J.~Auffenberg},
\author[Bartol]{X.~Bai},
\author[Madison]{M.~Baker},
% ...

\collab{(IceCube Collaboration)} 
\vspace{0.5cm} 

% ...
\address[Christchurch]{Dept.~of Physics and Astronomy, University of Canterbury, Private Bag 4800, Christchurch, New Zealand}
% ...
\address[Gent]{Dept.~of Subatomic and Radiation Physics, University of Gent, B-9000 Gent, Belgium}
% ...
\address[Madison]{Dept.~of Physics, University of Wisconsin, Madison, WI 53706, USA}
% ...
\address[Bartol]{Bartol Research Institute and Department of Physics and Astronomy, University of Delaware, Newark, DE 19716, USA}
\address[Oxford]{Dept.~of Physics, University of Oxford, 1 Keble Road, Oxford OX1 3NP, UK}
\address[RiverFalls]{Dept.~of Physics, University of Wisconsin, River Falls, WI 54022, USA}
% ...
\address[Wuppertal]{Dept.~of Physics, University of Wuppertal, D-42119 Wuppertal, Germany}
% ...
\thanks[Erlangen]{affiliated with Universit\"at Erlangen-N\"urnberg, Physikalisches Institut, D-91058, Erlangen, Germany}
\thanks[Bari]{on leave of absence from Universit\`a di Bari and Sezione INFN, Dipartimento di Fisica, I-70126, Bari, Italy}
\thanks[Goddard]{NASA Goddard Space Flight Center, Greenbelt, MD 20771, USA}
\corauth[Tosi]{Corresponding author.  Address: DESY, D-15735 Zeuthen, Germany, (delia.tosi@desy.de).} 

\newpage

\begin{abstract}
Using the South Pole Acoustic Test Setup (SPATS) and a retrievable transmitter deployed in holes drilled for the IceCube experiment, we have measured the attenuation of acoustic signals by South Pole ice at depths between 190 m and 500 m.  Three data sets, using different acoustic sources, have been analyzed and give consistent results.  The method with the smallest systematic uncertainties yields an amplitude attenuation coefficient $\alpha=3.20 \pm 0.57~\rm km^{-1}$ between 10 and 30~kHz, considerably larger than previous theoretical estimates.  Expressed as an attenuation length, the analyses give a consistent result for $\lambda \equiv 1/\alpha$ of $\sim$300~m with 20\,\% uncertainty.  No significant depth or frequency dependence has been found.
\end{abstract}


\end{frontmatter}



\section{Introduction}

Experiments to study ultra-high-energy neutrinos have been the subject of increasing interest during recent years \cite{AR05,AR06,AR08}. As observed by HiRes \cite{HI08} and the Pierre Auger Observatory \cite{AU10}, the charged cosmic ray flux decreases steeply above $10^{\,19.5}$\,eV. This is most likely due to the interactions of charged cosmic particles with the cosmic microwave background radiation, known as the GZK effect \cite{GE66,ZK66}. The detection of neutrinos from this interaction would confirm this explanation.  Spectral, temporal, and directional distributions of such neutrinos, enabled by detecting a significant number of them, would address important questions of cosmology, astrophysics and particle physics \cite{RI05}.

Estimates of the small flux of GZK neutrinos vary by an order of magnitude. The results from Engel, Seckel, and Stanev~\cite{ESS} are often used as a standard for the discussion of possible detector scenarios. In all cases, detector effective volumes of 100~km$^3$ or larger are required.

Several past and current experimental projects seek to detect GZK neutrinos. The best limits currently come from searches for radio signals from neutrino interactions. Presently the most stringent flux limit is from the ANITA project \cite{AN10}. However, all experiments searching for weak particle fluxes must contend with challenging systematic effects and background separation. This may be overcome in the future by using a hybrid detector. The radio technology could be complemented  by adding acoustic detectors, searching for the sound produced by the interactions of neutrinos above $10^{\,18}$\,eV \cite{AS79,LE79}. Simulations of a radio-acoustic hybrid detector encompassing the optical IceCube neutrino observatory at the South Pole \cite{IC06}, assuming an attenuation length on the order of a few kilometers as theorized in \cite{PR06}, gave a predicted detection rate of 20 neutrino events per year, half of them detected by both radio and acoustic sensors~\cite{BE05}.

Ice seems to be a favorable medium for the application of  the optical, radio and acoustic detection techniques. Ice properties have been measured for optical signals by the AMANDA experiment \cite{AC06} and for radio waves by the RICE experimental program \cite{RAT,RI06}. As far as the propagation of acoustic signals is concerned, theoretical estimates~\cite{PR06} have indicated low absorption and scattering corresponding to an attenuation length greater than 1~km.

To test the theoretical estimates, the South Pole Acoustic Test Setup (SPATS) was deployed in January and December 2007 in the shallowest 500 m of IceCube holes at the South Pole. SPATS is used to study the following four properties:
\begin{itemize}
\item the speed of sound as a function of depth in the ice, important for the expected neutrino signal strength, event localization and reconstruction;
\item	the acoustic noise level at the South Pole, determining the energy threshold of a future neutrino detector;
\item	the rate and nature of transient acoustic signals which could mimic neutrino interactions, presenting a serious background source;
\item the attenuation of acoustic signals by the ice, which determines the necessary density of acoustic sensors in the ice for a reasonable detection efficiency for neutrino interactions.
\end{itemize}
The sound speed has been measured in \cite{SPATS2}. Noise and transients are currently under study. Our measurement of attenuation by the ice is reported here.

\section{Experimental setup}
\label{secExperimentalSetup}

\subsection{The South Pole Acoustic Test Setup}
\label{secSPATS}
\begin{figure}[h]
 \centering
 \includegraphics[width=8.0cm]{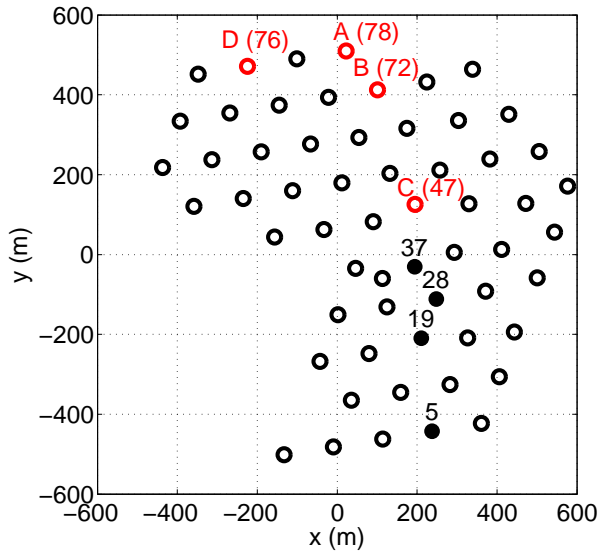} 
 \caption{SPATS geometry and IceCube footprint in 2009. The permanently deployed SPATS hardware is indicated by open circles with the string ID (ABCD) and corresponding IceCube hole number. The full circles indicate the location of the holes, indexed by the corresponding IceCube hole number, where the pinger was deployed in the 2008-2009 season.}
\label{fig:SpatsGeometry}
\end{figure}

Figure~\ref{fig:SpatsGeometry} shows the geometry of the acoustic instrumentation used for the measurement. In this section we provide a description of all the hardware used, while in the next we explain which sound source and which subset of sensors we used for each measurement.  The permanently deployed in-ice SPATS hardware consists of four vertical instrumented cables (\textit{strings}), which were deployed in the shallowest 500~m of four IceCube holes. The estimated surveying error on the position of each hole center is $\pm$0.1~m in each of the $x$ and $y$ coordinates.
 
The maximum string-to-string horizontal distance is 543~m between String~C and String~D. Strings A, B and C have \textit{acoustic stages} (described below) installed at 80, 100, 140, 190, 250, 320 and 400~m depth. String D has instrumentation at 140, 190, 250, 320, 400, 430 and 500~m depth. A diagram of the deployed instrumentation is presented in \cite{SPATS2}.

An acoustic stage consists of a transmitter module and a sensor module.  The transmitter module consists of a steel pressure vessel of $\sim$10.2 cm diameter that houses a high-voltage (HV) pulse generator board and a temperature or pressure sensor. The active element, a ring-shaped piezo-ceramic element cast in epoxy for electrical insulation, is positioned $\sim$13~cm below the steel housing. The HV pulse stimulates the piezo-ceramic element, resulting in an acoustic pulse. The amplitude of the acoustic pulse can be modified externally by adjusting two additional input signals: the steering voltage and the trigger pulse length. 

The SPATS sensor module has three channels, separated by 120$^{\,\circ}$ in azimuth to ensure good angular coverage. A channel consists of a cylindrical piezo-ceramic element that is directly soldered to a three-stage amplifier and pressed against the steel housing. The three channels of each sensor are independent and therefore can be treated as three stand-alone receivers.  From previous laboratory tests, it is known that the sensitivity of each sensor depends on both the polar and the azimuthal angles~\cite{Boeser07}. Most of the sensors were calibrated in water before installation, but there is evidence that the sensitivities have changed since deployment.  The effect on the sensor sensitivity of simultaneous low temperature, high pressure and ice coupling (compared to water coupling where calibration was performed in the laboratory) is unknown, as is the precise location of the $\sim$10.2~cm diameter sensor vessels in the 75~cm wide IceCube hole. During re-freezing of the holes, ice with bubbles and perhaps cracks is formed around the sensors, possibly leading to a strong but unknown angular dependence. A method to calibrate the sensors' beam pattern \emph{in situ} has not yet been developed.

The \textit{Hydrophone for Acoustic Detection at the South Pole} (HADES) was designed and constructed to offer an alternative sensor with a different dynamic range.  A ring-shaped piezo-ceramic element is connected to a two-stage differential amplifier placed inside the ring. The assembly is located outside of the housing of the SPATS sensor module and is coated with polyurethane plastic resin in order to protect the electronics from water and ice. Two HADES modules, each with one sensor channel, are installed in place of SPATS sensors on String D, at 190 m and 430 m depth.

We refer to each sensor channel by the string identifier letter (ABCD), the number of the stage (1-7, from shallowest to deepest) and the number of the channel (0-2). For example, AS6-0 indicates channel 0 of sensor module number 6 of String A.
 
Each of the acoustic stages is connected to an \textit{Acoustic Junction Box} (AJB) at the surface. This AJB is a strong aluminum box buried under $\sim$2~m of snow and contains a rugged embedded computer (\textit{String PC}) with the electronic components necessary to digitize locally all signals.  Each AJB is connected to a computer (the \textit{Master PC}), located in the IceCube Laboratory, which stores the data until it is transferred by satellite to the Northern hemisphere or copied locally to tape.  A GPS-based IRIG-B time code signal provides absolute time stamping.  For a detailed technical description of the SPATS permanent hardware and its development, see~\cite{SPATS2,Boeser07}.

\subsection{Retrievable pinger}
\label{secPinger}
In addition to the equipment deployed in the ice, a retrievable transmitter (\textit{pinger}) was  designed as a unique acoustic source to be operated in multiple water-filled IceCube holes, prior to IceCube string deployment. A previous version of the pinger was used in the 2007-2008 season to measure the sound speed profile. Here we describe the pinger as modified for the attenuation measurement in the 2008-2009 season, when it was deployed in IceCube holes 5, 19, 28 and 37 (Figure~\ref{fig:SpatsGeometry}).

The pinger is an autonomous transportable device consisting of an acoustic stage which is lowered into the water, and an auxiliary box, sitting on the surface, which provides power and a trigger signal. The box was connected to the acoustic stage during operation through a steel-armored four-conductor cable, about 2700~m long, spooled on a winch, used to lower and raise the pinger.

The acoustic stage consists of a custom designed high-voltage pulser circuit and of a spherical piezoelectric ceramic emitter\footnote{model ITC-1001} from the International Transducer Company (ITC). The Transmitting Voltage Response (TVR) specification provided by ITC has dominant components in the frequency range between 10 and 30~kHz, with a peak of 149 dB re ($\mu$Pa/V @ 1~m) at the resonance peak. The resonance frequency has been measured to be $f_{res} = 17.7$~kHz. 

\begin{figure}[h]	
\begin{center}
\noindent\includegraphics*[width=0.47\textwidth, height=0.37\textwidth]{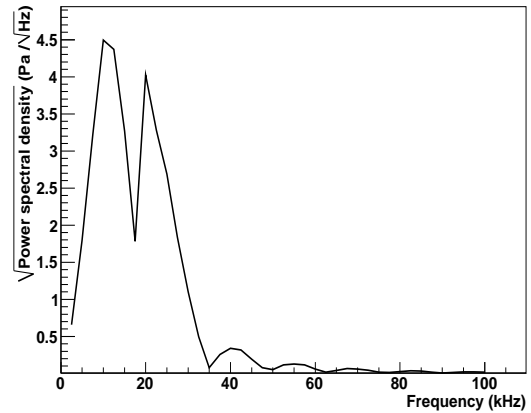}
\caption{\label{pinger_lin} Spectrum of the acoustic pinger pulse, calculated as the convolution of the electrical high-voltage pulse and the Transmitting Voltage Response of the \textit{ITC-1001}.}
\end{center}
\end{figure}

The box on the surface, (the \textit{Acoustic Pinger Box}, APB) contains a 24~V sealed lead acid rechargeable battery pack (Hawker-Cyclon) rated to low temperatures.  A GPS clock (Garmin model \textit{GPS 18 LVC}) generates a 1 pulse-per-second (PPS) signal. 
A frequency multiplier circuit, consisting of a Complex Programmable Logic Device (CPLD) and an oscillator, produces a continuous train of electric pulses at a rate which is adjustable by a switch. The rate of the pulses was 10~Hz (in holes 28, 19 5) or 8~Hz (in hole 37) during operation at the South Pole.  This means that a sequence of pulses equally spaced in time (by 100 or 125 ms) was produced by the CPLD, with every 10th (or 8th) pulse synchronized with the PPS rising edge.  Each electric pulse is delivered through the APB to the acoustic stage. Here it triggers a timer in monostable configuration which charges an LC circuit for a defined time, at the end of which the energy accumulated is transferred to the piezoelectric ceramic, resulting in acoustic emission.  The electric pulse exciting the piezoelectric ceramic is unipolar, about  300 V high, and has a full-width-half-maximum width of 60\,$\mu$s.  The pressure signal transmitted to the ice is the convolution of the stimulating electrical pulse and the TVR. The calculation of the spectrum, displayed in Figure~\ref{pinger_lin}, shows that most of the power is emitted in two lobes, one around 10 kHz and one around 20 kHz, with a minimum at the resonance frequency of the piezoelectric ceramic. 

The pinger was used in 2007-2008 to measure the sound speed vs. depth in South Pole ice for both longitudinal or pressure waves (P waves) and transversal or shear waves (S waves) \cite{SPATS2}. The production of shear waves (from mode conversion at the water-ice boundary) in that data set was favored by the fact that the pinger's lateral position in the hole was off-center and varying.  However, this made attenuation measurement difficult due to the change in shape of the waveform recorded by one sensor channel for different pinger holes. For the 2008-2009 season, the acoustic stage was equipped with mechanical centralizers, suitable to keep the acoustic emitter close to the central axis of the hole (see description in~\cite{Tosi10}). This prevented the stage from swinging and stabilized the acoustic pulse transmitted in the ice. The centralization of the pinger (and the generally longer distances between pinger and sensors in 2008-2009 than 2007-2008) minimized the appearance of shear waves in the 2008-2009 data set; consequently it has been possible to measure the attenuation of P waves without the complication of S waves. 

\section{Data processing and analysis techniques}

\subsection{Data sets}
\label{secDataProcessing}
Three different types of data were acquired and analyzed to measure the acoustic attenuation:
\\
\begin{enumerate}
\item Pinger data: multiple acoustic pulses were emitted from several water-filled IceCube holes and recorded by sensors of the SPATS array. For each pinger depth, 180 (144 in hole 37) single acoustic pulses were recorded by each sensor channel.
\item Inter-string data: transmitters of the SPATS array emitted acoustic pulses that were recorded by the SPATS sensors. For each transmitter 500 pulses were recorded by each single sensor channel.
\item Transient data: sound produced in re-freezing IceCube holes at depths of about 250~m was recorded by the sensors of the SPATS array.
\end{enumerate}
An overview of hardware components  involved in the analysis, together with the corresponding range of distances (baselines) and number of pulses recorded for each channel, is shown in Table~\ref{data_types}.

\begin{table}[hbtp]
\begin{minipage}[b]{0.47\textwidth}
\caption{\label{data_types}Overview of data used for the attenuation measurement.  The last column gives the number of pulses in each sensor recording.}
\resizebox{1\textwidth}{!}{
\begin{tabular}{p{0.2\textwidth} p{0.18\textwidth} p{0.22\textwidth} p{0.2\textwidth} r}
      Data type 	& Receiver 		& Source 		& Distance  & Number of pulses  
      \\
	\hline
	\hline
      Pinger  		& SPATS sensors 	& pinger    		& 125-1023~m &  144-180   \\ 
	\hline
      Inter-string    	& SPATS sensors     	& SPATS transmitters    & 125-686~m  & 500   \\ 
	\hline
      Transients      	& SPATS sensors     	& transient events    	& 243-750~m & 1   \\
      \hline
\end{tabular}
}
\end{minipage}
\end{table}

\subsection{Data processing}

For each of the analyses described below, $N_p$ consecutive acoustic pulses from one source (transmitter or pinger) are recorded by one sensor channel and are averaged to improve the signal-to-noise ratio.  Each string uses a single clock to drive both its analog-to-digital converters (ADCs) and its digital-to-analog converters (DACs). The ADCs are used to record the sensor waveforms with a sampling frequency of 200 kHz. The DACs are used to pulse SPATS transmitters. The clocks drift at a rate that is typically several parts per million, or tens of $\mu$s over the duration of a single sensor recording. This cumulative amount of drift is on the order of one signal oscillation period and therefore can cause severe decoherence in pulse averaging if the nominal rather than true sampling frequency is used.  On the sensor side we correct the clock drift effect by using the IRIG-B GPS signal (which is recorded synchronously with each sensor channel) to determine the actual sampling frequency at the time of the recording.  We then use this actual sampling frequency to average the recorded pulses.

The pinger pulse emission is driven by the GPS receiver and by the frequency multiplier. The GPS receiver provides the PPS pulse with a delay of about 1$\,\mu$s relative to the GPS signal which drives the String PC ADCs. The frequency multiplier synchronizes the train of electric pulses with the PPS pulse, but introduces some jitter in time which we measured in the laboratory to be $\pm\,5\,\mu$s over the recorded duration for a single channel.  Time jitter smears out the amplitudes contributing to each single point in the averaged sensor waveform, the effect of which is automatically included in the statistical uncertainty of the average pulse amplitude.

Inter-string data (recorded with the frozen-in transmitters rather than the retrievable pinger) have the additional complication that the transmitters are not synchronized to GPS.  These transmitters pulse at a rate which can drift relative to absolute time.  Therefore in this data, in addition to the drift of the recording sensor, we must correct for the clock drift on the transmitting string, which causes the actual transmitter repetition rate to be different from the nominal one. For each run we measured the actual repetition rate using IRIG-B GPS signals in the same way as described above for the sensors. The true repetition rate varied from string to string but was constant on each string at the few-percent level over the course of the two-day inter-string data taking campaign. Therefore for each string we used the mean value of the transmitter repetition rate in averaging the recorded pulses.  The effect of residual clock drift due to using the average drift rate at each transmitting string rather than using the instantaneous drift rate is $< 0.3$~sample over the duration of a 20~s sensor recording, much less than a single signal oscillation period and therefore not sufficient to cause any residual decoherence in pulse averaging.

After correcting the sample times for clock drift, we wrap the waveform samples in time modulo the pulse repetition period, in order to overlay all $N_p$ pulses recorded in a single sensor recording. We then time-order the samples and bin them in groups of $N_{bin}$ consecutive samples. The number of samples per bin $N_{bin}$ was chosen to be one-half the number of pulses for the inter-string data, and equal to the number of pulses for the pinger data.  The choice of bin size was made independently for the inter-string and pinger data because they have different frequency content and signal-to-noise ratio.  For each bin $j$, we compute the mean voltage $V_j$ and the standard error of the mean $\sigma_{V_j}$, which we use as an estimate of the uncertainty on the sample amplitude because the noise is observed to be stable and Gaussian. The resulting average pulse is one repetition period in duration.  For a detailed description of the algorithm used see \cite{Vandenbroucke09,Tosi10}.

\subsection{Analysis techniques}
\label{secAnalysisTechniques}

The basic quantity measured by the sensors is a pressure wave exciting their piezo-ceramics for a duration which depends on the sensor construction, the signal strength and the distance between emitter and receiver (see Figures~\ref{waveforms_consistency:1} and \ref{waveforms_consistency:2}). The signal amplitude $A$ recorded by a sensor is proportional to the input acoustic pressure. For a point source with spherical emission, $A$ depends on the distance to the source and the amount of attenuation by the ice:

\begin{equation}
A  =  \frac {A_0} {d} e^{-\alpha d} = \frac {A_0} {d} e^{-d/\lambda}. 
\label{def_point_amp}
\end{equation}
where $A_0$ is constant for a given transmitter and sensor channel and depends on the amplitude of the sound emitted at the source and the sensitivity of the receiver; $d$ is the distance to the source; $\alpha$ is the acoustic attenuation coefficient and the attenuation length $\lambda$ is its inverse.  We assume here that $\alpha$ is independent of frequency, position, and direction.  We multiply both sides of Equation~\ref{def_point_amp} by the known distance $d$ and take the natural logarithm to define a new variable $y$:

\begin{equation}
y = \ln(A d)
\label{y_definition}
\end{equation}

\noindent In this way we can turn the previous nonlinear equation into a linear one:

\begin{equation}
y = \ln A_0 -\alpha d =  -\alpha d + b,
\label{fit_equation}
\end{equation}

\noindent where $b$ is a free normalization parameter related to the sensitivity of the particular sensor piezoceramic. Linear regression can be applied to determine the best fit and uncertainty on each of the two parameters $\alpha$ and $b$. 

The measured signal has a complex waveform that depends on the particular choice of transmitter and sensor (see Figures~\ref{waveforms_consistency:1} and \ref{waveforms_consistency:2}).  It is difficult to use one particular peak amplitude of this waveform  directly to characterize an arriving signal. A more stable quantity is the \textit{energy}  of the recorded pulse,

\begin{equation}
E \sim\sum_{i=1}^n V_i^2,
\label{energy_definition}
\end{equation}

\noindent where $V_i$ is 
the amplitude of the sample $i$ (after averaging, as explained in section \ref{secDataProcessing}) and the number of samples $n$ is waveform-dependent. Because every sample $i$ of the mean waveform has a statistical uncertainty $\sigma_{V_i}$, we can calculate a statistical uncertainty $\sigma_E$ for the energy.  We note that this quantity is an ``energy'' in the signal-processing sense and is directly proportional to the acoustic energy in the pressure pulse.

From the energy $E$, once the noise has been subtracted as explained below,  we can also calculate the \textit{effective amplitude} of the pulse:

\begin{equation}
A_{eff} = \sqrt{E}
\label{effective_amplitude}
\end{equation}
\noindent
and thereby extract the amplitude attenuation coefficient from Equation~\ref{fit_equation} (using $A = A_{eff}$) rather than the energy attenuation coefficient. The statistical uncertainty of the effective amplitude is determined with standard error propagation from $\sigma_{E}$.  This method has been used for most of the studies described below. 

An alternative approach is to apply Eqs.~\ref{def_point_amp} - \ref{fit_equation} to a calculation of the waveform energy in the frequency domain. In this case, the effective amplitude $A_{eff}$ is given by

\begin{equation}
A_{eff} = \sqrt{ \sum _m  \left| \hat S_m \right|^2 }, 
\label{amplitude_fdomain}
\end{equation}
\noindent
with the coefficients of the noise-subtracted Fourier spectrum $\hat S_m$.  Results of both methods will be given below.

In both methods, we do not deal directly with the response function of the sensors, which is implicitly included in $A_0$. The frequency-dependent sensitivity of our sensors exhibits several resonances due to both the piezoelectric ceramics and the housing, with their corresponding couplings. As discussed in section \ref{secDataAnalysis}, in the pinger measurements we assume that the sensor response function is constant over time for all waveforms recorded with the same sensor channel (therefore Equations~\ref{def_point_amp}-\ref{fit_equation} can be applied exactly since $A_0$ is really independent of $d$).  In the other analyses (inter-string and transient data) the variation of $A_0$ is included as a source of systematic uncertainty.

\section{Pinger measurement}
\label{secPingerAnalysis}

\subsection{Pinger data acquisition}
\label{secDAQ}
As described in Section~\ref{secPinger}, the pinger was operated in four IceCube holes (28,~19,~5,~37), shown in Figure~\ref{fig:SpatsGeometry}, in the 2008-2009 season. The pulse repetition rate was 10~Hz in all holes except for hole 37, for which the repetition rate was 8~Hz. The distances from the SPATS array ranged from 156~m (String C to Hole 37) to 1023~m (String D to Hole 5).  The range in azimuth angles spanned by the pinger holes as seen from an acoustic sensor was small: 13~degrees for String D, 7.2 degrees for String A, 6.6 degrees for String B, and 8.2 degrees for String C.

The pinger was lowered from the surface to a maximum depth of 500~m and then raised back to the surface. Both on the way down and on the way up, the pinger was stopped for $\sim$5~min at nominal depths of 190, 250, 320, 400, 500~m, which are instrumented with SPATS sensors. The depth was monitored using the cable payout, initially calibrated by counting the number of turns of the winch. In addition, multiple calibrated pressure sensors attached to the acoustic stage recorded the hydrostatic pressure as a function of time. After deployment we averaged the pressure and the payout data and verified that the stop depths were within $\pm\,$5~m of the nominal values.

Only those recordings of sensors at the same depth of each pinger stop depth were analyzed for attenuation, in order to perform the measurement along horizontal paths (see more details in the section \ref{secDataAnalysis}); all three channels of the same sensor were recorded simultaneously (differently from \cite{SPATS2}).

Due to the time necessary to transfer the data from the String PC to the Master PC through a DSL connection over a surface cable, the optimal duration of the recording at 200~kHz on the 3 channels of one sensor was found to be 18~s and this duration was used for the pinger runs.

\subfiglabelskip=0pt
\begin{figure}[h]	
\begin{center}
\subfigure[][]{
\label{waveforms_consistency:1}
\noindent\includegraphics*[width=0.47\textwidth, height=0.37\textwidth]{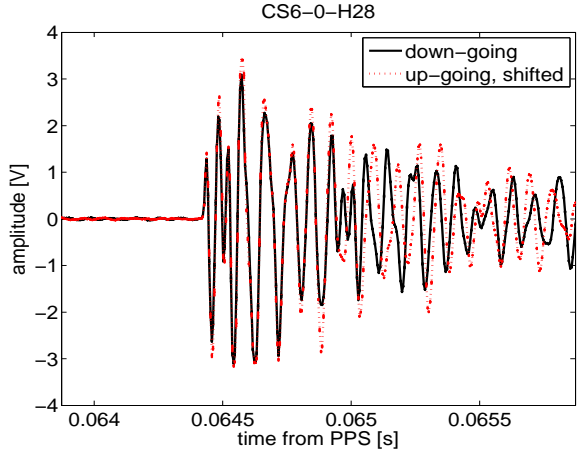}}
\subfigure[][]{
\label{waveforms_consistency:2}
\noindent\includegraphics*[width=0.47\textwidth, height=0.37\textwidth]{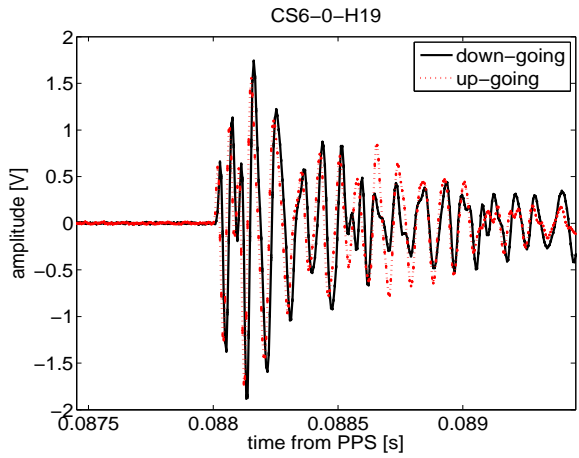}}
\caption{\label{waveforms_consistency}Waveforms recorded by the same channel when the pinger was stopped while lowering and while raising, for two different holes at a distance of 243~m (top) and 336~m (bottom). Clock drift correction and averaging of the 180 recorded pulses have been applied. In the top plot, one of the waveforms has been shifted in time to overlap with the other one.  This small shift is due to a slight difference in the pinger depth between the two runs (see \cite{Tosi10}).}
\end{center}
\end{figure}

As can be seen in Figure~\ref{waveforms_consistency}\subref{waveforms_consistency:1}, two waveforms recorded while the pinger was stopped during lowering and raising, respectively, within the same hole look very similar and overlap very well.  This is thanks to the mechanical stabilization of the pinger in the hole, in contrast to the 2007-2008 data set where the pinger was freely swinging.  The comparison of these waveforms with those of Figure~\ref{waveforms_consistency}\subref{waveforms_consistency:2} shows that waveforms recorded by the same channel are very similar even when the pinger is moved to a different hole.

\subsection{Pinger data analysis}
\label{secDataAnalysis}
For all analyses described below, we selected data recorded by each channel when the pinger was at the same depth as the sensor, in every instrumented hole. This selection provides 49 combinations (four times three SPATS channels on each of the four strings, plus one HADES channel) which can be used for the same number of independent attenuation measurements.
This data set provides the most systematic-free attenuation analysis, because it allows us to compare recordings of the same sensor as the transmitter is moved among holes at varying distances from the sensor.  Moreover, the sensor-transmitter configurations that were selected feature minimal variation in both polar angle and azimuthal angle of the arriving signal at each sensor, as the pinger is moved from hole to hole. We emphasize that for each single measurement only data recorded by one single channel have been used. This means that we are not sensitive to variation of sensitivity between channels and we can neglect the sensor response function, unknown in our case, assuming the following:
\begin{itemize}
\item The sensor response is constant throughout the pinger data taking.  This is supported by the fact that the noise spectra measured by the sensors has been demonstrated to be very stable in time.
\item The sensor output is linear in amplitude with respect to the input amplitude.  This has been demonstrated in the laboratory for signals which are within the range of amplitudes considered here. Waveforms which are saturated are excluded.
\end{itemize}
We also make two additional important assumptions, which are supported by the similarity of the waveforms recorded by each channel for multiple pinger distances and by the analysis of the received signal in the frequency domain (see section \ref{freqDomainPingerAnalysis} and in particular Fig. \ref{FD_spectrenx2}):
\begin{itemize}
\item The pinger emission is constant throughout all sets of measurements taken.
\item The sound transmission in the medium is not affected by dispersion, \textit{i.e.}, the frequency content of the pulse is independent of the traveled distance. 
\end{itemize}

Two types of analyses have been performed on the pinger data: the energy of the full waveform calculated in the time domain and the energy calculated in the frequency domain. 

\subsubsection{Time domain pinger analysis}

The energy analysis integrates over the complete waveform. The full waveform recorded when the pinger was stopped at the depths of the sensors was processed as explained in Section~\ref{secDataProcessing}. The high quality averaged pulse obtained was used to calculate the energy $E_{S+N}$ for each channel-hole combination by applying Equation~\ref{energy_definition}. This includes both a signal and a noise contribution. The noise energy $E_N$ is calculated using the data recorded immediately before and after the pinger operation, which was verified to be very stable over the time. The processing of the waveform is done in the same way for both kinds of data (signal plus noise or noise-only). The noise was subtracted to estimate the signal energy $E = E_{S+N} - E_N$. Next, the effective amplitude was calculated (Equation~\ref{effective_amplitude}). The number of valid measurements was 48 of the 49 channels available, as one channel (HADES, at level 190~m in string D) did not have more than two data points which pass the cut $E>0$.

The statistical error on the effective amplitude was determined by error propagation from the statistical uncertainty of the sample amplitudes  as explained in section \ref{secAnalysisTechniques}. To determine the systematic uncertainty, we calculated for each of the 48 measurements the $\chi^2/n_d$ (where $n_d$ is the number of degrees of freedom of the fit) for different hypothesized systematic error values and we observed how the distribution of these values changed. We found that adding 15\% systematic uncertainty to each fit brings this distribution to have mean equal to 1. We therefore used this value as an estimate of the systematic uncertainties among data points for each fit, which can be attributed to residual azimuthal and polar variation of the sensor sensitivity (due to a non-perfect alignment of holes and stopping depths).  We also found that variation of the assumed systematic uncertainty within a reasonable range does not significantly affect the final result (see \cite{Tosi10}).

\begin{figure}[tbp]	
\noindent\includegraphics[width=.45\textwidth]{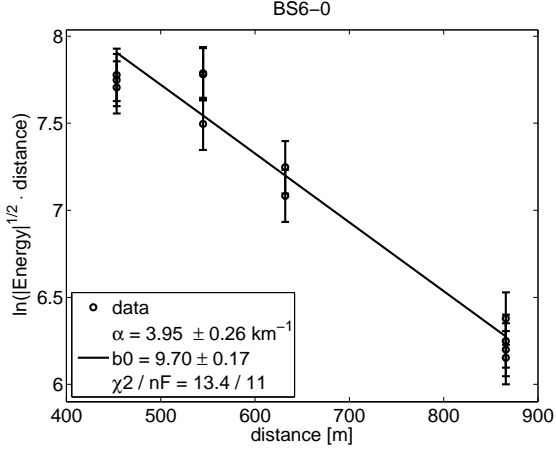}
\caption{Example fit of effective amplitude vs distance for sensor channel BS6-0 installed at 320~m depth. Error bars include the statistical and systematic uncertainty, estimated as explained in the text.}
\label{energy_fit_1_channel}
\end{figure}

Taking into account the statistical and systematic uncertainties, we fit Equation~\ref{fit_equation} to determine the amplitude attenuation coefficient $\alpha$. A typical fit of the data, from sensor channel BS6-0, is shown in Figure~\ref{energy_fit_1_channel}.

\begin{figure}[tbp]
\noindent\includegraphics[width=.45\textwidth]{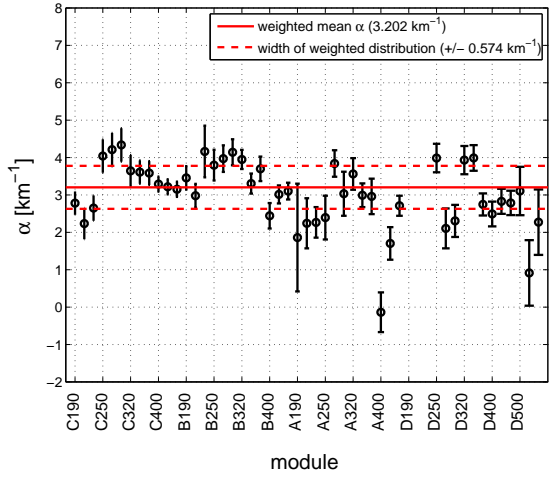}
\caption{Attenuation coefficients (with standard error) obtained by energy analysis in time domain. The three channels of each module are indicated by the three data points at and to the right of the corresponding label on the horizontal axis.}
\label{energy_all_alphas}
\end{figure}

A compilation of results obtained by fitting the single-channel data is shown in Figure~\ref{energy_all_alphas}. The strings have been sorted from nearest to farthest from the pinger holes and, within each string, modules have been sorted by depth. 
For details of each single-channel fit see \cite{Tosi10}. A consideration must be made looking at the distribution of the points: the scatter of the data, greater than the error bars, implies that there are additional systematic uncertainties (for example, local properties of the ice, or of the interface between hole ice and sensors) which we are not able to identify and quantify without looking at the spread of the data. Nevertheless, the 48 measurements allow us to constrain the attenuation coefficient within a narrow range.
To take into account the uncertainties of the individual measurements we assign to each entry a weight inversely proportional to the error of the value:
\begin{equation}
w_{i} = \frac{\frac{1}{\sigma_{\alpha,i}^2}}{\sum_i \frac{1}{\sigma_{\alpha,i}^2}}
\label{eq:weights}
\end{equation}
where $\sigma_{\alpha,i}$ is the one-sigma uncertainty on the value of $\alpha$ in the measurement $i$ obtained from the fit.  
We then build a histogram using all the weighted entries and we take the mean and the standard deviation (indicated by the dashed line in Figure~\ref{energy_all_alphas}) as final value and error:

\begin{equation}
\left<\alpha\right> = 3.20 \pm 0.57~{\rm km^{-1}}.
\end{equation}

\begin{figure}[tbp]	
\noindent\includegraphics[width=.45\textwidth]{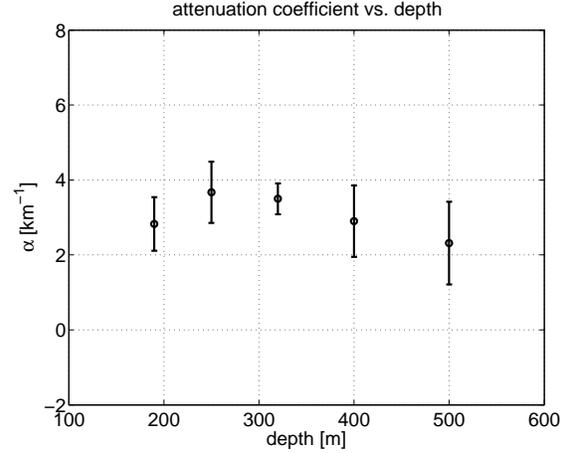}
\caption{Mean attenuation coefficient as a function of the depth.}
\label{energy_all_depths}
\end{figure}

In Figure~\ref {energy_all_depths} the average attenuation coefficient is shown for each depth. No clear depth dependence is observed.

\subsubsection{Frequency domain pinger analysis}
\label{freqDomainPingerAnalysis}
The data selected for the frequency domain analysis are the same as in the previous analysis, \textit{i.e.} the attenuation coefficient is obtained from the horizontal pinger-sensor configuration. However, the differences in the method, including averaging and background handling, allow for a cross-check of the methodology.  Moreover the analysis software used for the two methods was completely separate, allowing for an independent cross-check.

For each channel a series of pinger pulses ($l = 1, ...,  N_p$; $N_p =$180 or 144) is recorded. Each pulse is the sum of signal $s$ and an additive stationary noise component $n$ (uncorrelated with the signal), therefore the amplitude $x$ of a sample $k$ at time $t_k$ can be written as:   

\begin{equation}
x ^l (t_k)  =  s ^l (t_k) + n ^l (t_k).
\end{equation}

The discrete Fourier transform yields the complex numbers

\begin{eqnarray}
X^l (\omega_m) & = & S^l (\omega_m) + N^l (\omega_m ) \\
& = & \frac{1}{\sqrt{M}} \, \sum_{k = 0}^ M x_k ^{l} \, {\rm e} ^{-i \omega_m t_k  },
\end{eqnarray}

\noindent where ($k = 1, ..., M$) is the sample number within each pinger pulse ($M = 20$ or 25).

\begin{figure}[tbp]
\noindent\includegraphics[width=20pc]{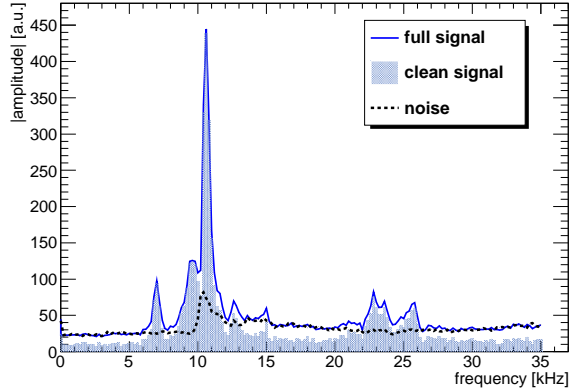}
\caption{Example of frequency spectra of signal plus noise, pure noise and signal after noise subtraction }
\label{examplenoiseSub-H37-AS5-1}
\end{figure}

The average noise, $\hat N$, is estimated from the off-pulse portion of the waveforms (according to the methods described in \cite{1163209,1170788,QualityImp}) by averaging over a set of pure noise samples taken from each waveform.
To avoid any overlap with the signal, the noise intervals are taken before each recorded pulse. Examples of a raw signal plus noise spectrum, a pure noise spectrum and a signal spectrum are shown in Figure~\ref{examplenoiseSub-H37-AS5-1}.

\begin{figure}[tbp]
\noindent\includegraphics[width=20pc]{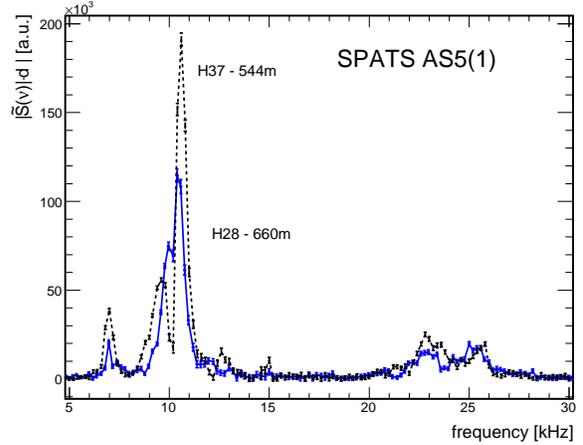}
\caption{Example of typical noise-subtracted spectra. The two spectra are taken from the same sensor, but for different distances. The Fourier magnitudes are multiplied by distance, which then allows a direct estimate of the attenuation, as described by Equation~\ref{fit_equation}.}
\label{FD_spectrenx2}
\end{figure}

As can be seen from Figure~\ref{FD_spectrenx2}, the spectral shape is approximately constant for the same sensor, but attenuated with increasing distance, i.e. for different pinger hole measurements.  The peak at 10~kHz reflects a characteristic peak in the sensor response.

We compute the average signal spectrum density $\hat S_m$ after subtracting the noise density from each pinger pulse:

\begin{eqnarray}
\left| \hat	S_m  \right|^2 &=&  \frac{1}{N} \sum_l \left[ \left| X_m ^l \right|^2- \left| \hat	N_m \right|^2 \right]  
\end{eqnarray}

\noindent where the sum extends over the pulses and applies in the region 5 - 30 kHz. Next  we compute the effective amplitude $A_{eff}$ via Equation~\ref{amplitude_fdomain} and apply the same fit procedure as in the time domain analysis.

The weighted mean of $\alpha$ for the available channels from the frequency domain analysis is
\begin{equation}
\left< \alpha \right> = 3.75 \pm  0.61~{\rm km}^{-1},
\end{equation}

\noindent where the uncertainty is given by the width of the weighted distribution. The difference of the value from that of the time domain analysis is explained by the different method used to separate signal from noise; the two results are nevertheless compatible. 

A similar study has been done using as input the averaged waveforms, separating the energy contribution for the frequency intervals between 5 and 17 kHz, between 17 and 35 kHz, and for the whole interval 5 to 35 kHz.  We investigated the possible presence of a trend in the variation of $\alpha$ over each interval compared to the one obtained using the full spectrum, and we have not found any significant frequency dependence. In some cases the attenuation below 20~kHz is stronger than around 30~kHz (as seen for example in Figure~\ref{FD_spectrenx2}); in other cases the variation is opposite. In all cases, the obtained $\alpha$ values agree within uncertainties~\cite{Tosi10}.  

Both SPATS and HADES sensors obtain the same shape for the frequency distribution, despite their different frequency responses.  The recorded spectrum in both cases is dominated by the pinger frequency content rather than by the response of the sensors.

\section{Inter-string measurement}
\label{secInterString}
In addition to the measurement performed with the retrievable pinger operated in water, we measured the acoustic attenuation using the frozen-in SPATS transmitters. Two independent analyses of a single data set were performed. The processing techniques used to average the waveforms and estimate the noise-subtracted signal energies use the same algorithms, in the time domain (see Section~\ref{secDataProcessing}) but were implemented independently to cross-check one another. The signal energies determined by the two analyses are consistent.

\subsection{Inter-string data acquisition and processing}
\label{secDAQAndProcessing}
Data were taken for each transmitter recorded by each sensor in the SPATS array.  In each run, one transmitter was pulsed at a 25~Hz repetition rate for 40~s. Data from all inter-string transmitter-sensor combinations were collected over a two-day period (April 1-2, 2009).  The transmitters require several seconds to reach steady pulse-to-pulse performance, likely due to self-heating of the electronics during initial operation followed by temperature equilibration.  The initial acoustic pulse amplitude is somewhat larger than in the steady state, followed by decay to the steady-state amplitude with a time constant of $\sim$2~s.  To be sure we are recording in the steady state, we start the sensor recording 11~s after the transmitter begins pulsing.  In each run, all three channels of a single sensor module are recorded continuously for 20~s, enough time to record 500 transmitter pulses.  The three channels are sampled synchronously, at 200 kilosamples per second on each channel.  

The ambient acoustic noise conditions at the South Pole have been determined by SPATS to be very stable~\cite{Karg09}.  In particular we checked that they were stable during the two-day inter-string data-taking period.  Several raw noise runs were taken on each channel, interspersed among the transmitter recordings, using the same sampling frequency and time duration but with no transmitter pulsing.  The noise runs were processed with the same waveform averaging algorithm as the transmitter runs.  From this data the DC offset ($\mu$) and the standard deviation of the noise samples ($\sigma$) were calculated for each sensor channel.  Each of $\mu$ and $\sigma$, on each sensor channel, was stable at the few-percent level during the two-day period.

For each sensor recording, the mean waveform was processed to determine the noise-subtracted signal energy following the algorithm described in Section~\ref{secDataProcessing}.

\subsection{Inter-string single-depth direct analysis}
\label{secDirect}
This analysis uses a single transmitter recorded by all sensor channels at the same depth as the transmitter.  A single transmitter is used because the different transmitters are known to have different inherent transmittivity and perhaps different coupling to the ice with respect to one another.  Only those sensor channels at the same depth as the transmitter were used in order to mitigate effects due to the unknown change of transmittivity with varying polar angle.

For each transmitter there are typically three sensor modules, each with three channels, located at the same depth as the transmitter but on other strings.  The acoustic attenuation was measured for each transmitter in the array, after applying the following two quality cuts to the data.  First, the statistical uncertainty of the effective amplitude at a given channel was required to be 20\% or smaller, in order to be considered a ``good'' channel and included in the fit.  Second, there had to be at least one good channel at each of at least two distances from the transmitter, at the same depth as the transmitter, in order for a fit to be performed for a given transmitter.  Note that only five of the nine instrumented depths are instrumented on all four strings.  Twelve of the 28 SPATS transmitters met these two selection criteria.

For each good sensor channel, the quantity $y$ (defined by Equation \ref{y_definition}) was determined following the procedure described in Section~\ref{secDataAnalysis}.  We performed a linear fit for each of the 12 transmitters to determine the  acoustic attenuation coefficient $\alpha$ and the free normalization parameter $b$ related to the emission strength of the transmitter.

Both statistical and systematic uncertainties were calculated for each $y$ value for each channel recording each transmitter. The statistical contribution was determined by propagating errors from the statistical uncertainty of the effective amplitude. The systematic uncertainty is dominated by the unknown relative sensitivity of the sensor channels. We estimate the channel-to-channel variation in sensitivity by treating the pinger results as an \textit{in situ} calibration of the sensor channels. Since the pinger analysis fits the $b$ parameter for each channel independently, $e^b$ can be taken as a measure of the sensitivity of each channel.  
The results of this \textit{in situ} calibration were used to estimate the systematic uncertainty in $y$ for the inter-string attenuation analysis, which combines different sensor channels for each fit.  The absolute systematic uncertainty in $y$ was 0.97.  This was added in quadrature to the statistical uncertainty of $y$ for each channel.

\begin{figure}[tbp]
\noindent\includegraphics[width=20pc]{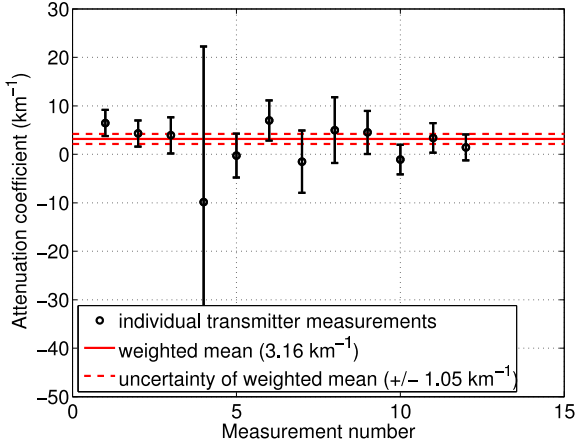}
\caption{Compilation of attenuation fits for the 12 transmitters used in the single-depth inter-string analysis.   The weighted mean is shown as a solid horizontal line, and the one-sigma uncertainty of the global fit is indicated with dashed lines.} 
\label{attenuation_coefficient}
\end{figure}

Results of the fits for the 12 transmitters are compiled in Figure~\ref{attenuation_coefficient}.  For each transmitter $i$, the best fit attenuation coefficient $\alpha_i$ and corresponding uncertainty $\sigma_i$ are shown.  Many of the single-transmitter fits individually give a result consistent with zero attenuation.  However, the 12 results can be combined to improve the precision of the measurement.  Determining the weighted mean of the measurements (weighting by $1/\sigma_i^2$) gives a global best fit of
\begin{equation}
\left<\alpha\right> = 3.16~\pm 1.05~{\rm km^{-1}}.
\end{equation}

\subsection{Inter-string multi-depth ratio analysis }
\label{secRatio}
None of the permanently deployed transmitters or sensors has been calibrated in ice.  Therefore, both an unknown inherent sensitivity $\mathcal{S}_{j}$ and transmittivity $\mathcal{T}_{i}$ enter into the equation for a single inter-string amplitude measurement. With these variables, Equation~\ref{def_point_amp} can be written for the combination of transmitter $i$ and sensor $j$ as:
\begin{equation}
A_{ij} =\frac{\mathcal{T}_{i}\mathcal{S}_{j}}{d_{ij}}e^{-\alpha d_{ij}}, 
\end{equation}
\noindent where $A_{ij}$ is the recorded amplitude of the pulse transmitted by transmitter $i$ as detected by sensor $j$ and $d_{ij}$ is the distance between transmitter $i$ and sensor $j$. If we then take two transmitter-sensor pairs, transmitters $T_{i}$ and $T_{k}$ heard both by sensors $S_{j}$ and $S_{l}$,  it is possible to construct a ratio~\cite{Boeser07} of amplitudes:
\begin{equation}
\ln(\mathcal{R}_{A}\mathcal{R}_{d})=\ln\left(\frac{A_{ij}A_{kl}}{A_{il}A_{kj}}\frac{d_{ij}d_{kl}}{d_{il}d_{kj}} \right)=-\alpha D_{x} + b, 
\label{ratio_eq}
\end{equation}
where $\mathcal{R}_{A}$ and $\mathcal{R}_{d}$ are ratios of amplitudes and distances respectively, $D_{x}=([d_{ij}-d_{il}]-[d_{kj}-d_{kl}])$ and $b$ is a free fit parameter introduced to allow for a systematic shift in $y$. All three channels of both sensor modules are required to have $A > 1.5 \sigma_{\mathrm{stat}}(A)$, where $\sigma_{\mathrm{stat}}(A)$ is the statistical error on the amplitude $A$. Dead and saturated channels are excluded.

One single measurement would yield the attenuation coefficient $\alpha$ if the transmitters and sensors were all perfectly isotropic. However, the azimuthal orientation of each sensor module is unknown and the transmitter signal is known to vary significantly with polar angle. Both effects need to be minimized and need to be accounted for. Therefore, for the analysis presented here, only transmitter-sensor combinations from the lower neighboring levels were used: (190, 250), (250, 320), and (320, 400)~m depth. This limits the difference in polar angle for an amplitude ratio to a maximum of 32$^{\,\circ}$. The maximum difference in azimuth angle is $\sim$120$^{\,\circ}$. An average sensor-module response is used so that the azimuth effect is less significant. 
A total uncertainty on the single amplitude ratio of 100$\%$ has been estimated by studying the spread of the $b$ values obtained from the pinger analysis, as described in Section \ref{secDirect}. On top of that, the variation due to the angular variation of the SPATS transmitters needs to be included. This was studied in the laboratory, where a maximum amplitude variation of 40$\%$ in transmitter emission was observed for the inter-string geometry discussed here. For the ratios with amplitudes at minimal polar angles, the assumed uncertainty on a single ratio is conservative. 


\begin{figure}[h]
\noindent\includegraphics*[width=18pc]{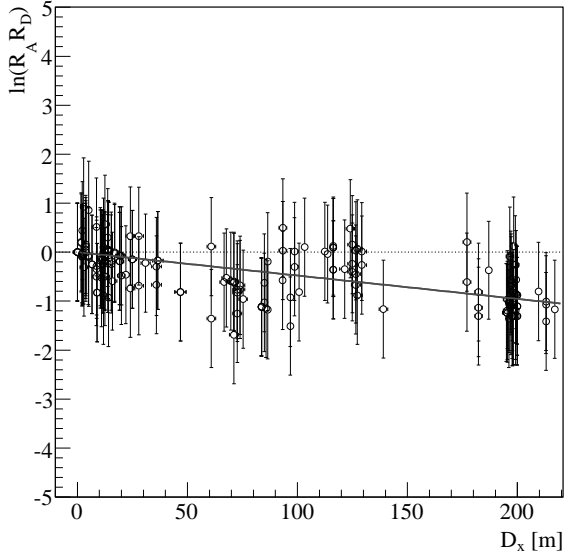}
\caption{The 2-level inter-string ratios as a function of $D_{x}$. The spread on the points is an indication of the residual polar and azimuthal systematic dependencies. The solid line shows the best fit; the dotted line shows $y=0$ for reference.} 
\label{fig:interstring_ratio}
\end{figure}

Figure~\ref{fig:interstring_ratio} shows all 172 ratios with the estimated systematic error bars and the best fit obtained by constraining the offset (\textit{i.e.} the parameter $b$ of Equation~\ref{ratio_eq}) to be 0. As mentioned above, the figure reflects the fact that the systematic errors are overestimated. However this does not influence the central value of the linear fit coefficient $\alpha$.  The fit results in a value
\begin{equation}
\alpha = 4.77 \pm 0.67~{\rm km^{-1}}.
\end{equation}

\section{Transients measurement}
\label{secTransients}
In addition to the signals from the retrievable pinger and the frozen-in transmitters, we can use ambient transient events to estimate the attenuation coefficient.  SPATS runs in a transient data acquisition mode during most of each hour.  Three channels from each of the four strings are monitored for large-amplitude events. A simple 5$\sigma$ threshold trigger is used, where $\sigma$ is the Gaussian noise amplitude.  The Gaussian noise is determined on each channel and accordingly a different absolute threshold is used on each channel.  Events above thresholds are recorded with an absolute GPS time stamp; this occurs with a frequency of $\sim$1~Hz per channel. The hits from the four strings (where a \emph{hit} is defined to be a timestamped waveform recorded from an individual channel) are time ordered offline and processed through an algorithm to find hits within a coincidence time window of 200 ms (the time necessary for a pressure wave signal to cross the SPATS array).
Each cluster of hits occurring within the coincidence time window is an \emph{event}. For events with more than four hits and at least three strings, a vertex location is calculated.

\begin{figure}[h]
\noindent\includegraphics[width=18pc]{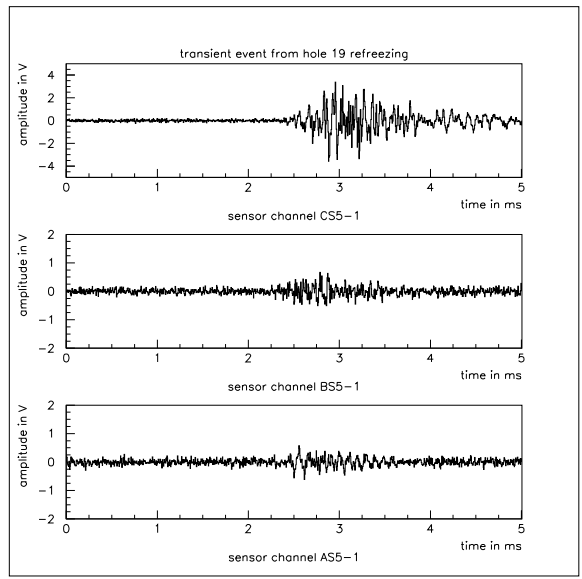}
\caption{Waveforms produced by sound from refreezing ice in  Hole 19 received at the  250 m sensors of Strings C, B and A at a distance of 336 m, 632 m, and 744 m respectively. The time scale refers to the trigger time which is independent
for each channel.} 
\label{hole19-12242008}
\end{figure}

Using this data taking mode for 45 minutes out of every hour, data have been collected since the end of August 2008. Acoustic signals have been observed for several days from re-freezing IceCube holes with precisely known x-y coordinates. Because some of these holes were also used for collecting pinger data, the corresponding sensor calibration constants calculated from those data can be applied to signals whose source is localized to near the corresponding pinger stop depths. Thirteen appropriate events have been found with source location at the coordinates of holes 19, 20 and 28, and depths between 230 and 270~m, which fit in the dynamic range of the SPATS sensors at all strings involved. The waveforms of an example event are shown in Figure~\ref{hole19-12242008}. Single-event analysis using effective amplitudes derived from energy in the time domain (see Section~\ref{secDataAnalysis}) leads to a mean attenuation coefficient 

\begin{equation}
\alpha = 3.64 \pm 0.29~{\rm km^{-1}},
\end{equation}

\noindent compatible with the results from both the retrievable pinger and the frozen-in transmitters. The relatively small errors in comparison to the pinger analyses may be due to the fact that here only sensors installed at one depth (250 m) have been used. The error shown is the statistical error only and neglects any additional uncertainty of the sensor calibration in comparison to the pinger analysis. The frequency spectrum of the transient events needs further study but contains contributions up to 80 kHz.

\section{Discussion and outlook}
\label{secDiscussion}

The results from the various analyses described above are summarized in Table~\ref{summaryTable}.

\begin{table}[hbtp]
\caption{\label{summaryTable} Summary of results from various attenuation analyses.}
\begin{center}
\begin{tabular}{l c c c}
      Analysis & Measurements & $\alpha$ [km$^{-1}$]  \\
	\hline
	\hline
      Pinger (time dom.)     & 48     & 3.20 $\pm$ 0.57    \\ 
	\hline
      Pinger (freq. dom.)    & 39     & 3.75 $\pm$ 0.61    \\ 
	\hline
      Inter-string direct    & 12     & 3.16 $\pm$ 1.05    \\  
	\hline
      Inter-string ratio       & 1    & 4.77 $\pm$ 0.67    \\  
	\hline  
      Transients               & 13     & 3.64 $\pm$ 0.29   \\
      \hline
\end{tabular}
\end{center}
\end{table}

All of the analyses yield consistent results. However, we cannot simply average them, due to the quite different systematic effects (as well as correlations among them) which affect each of them as discussed in detail in previous sections. In Table~\ref{systematicsTable} the different effects are summarized and attributed to the individual studies they affect. The pinger approach uses the same sensor and the same transmitter at the same zenith angle and nearly the same azimuthal angle at different distances, thereby mitigating all effects connected with unknown sensitivities of emitter and receiver. For this reason we believe it provides the most reliable result. Relying on the same data set, the pinger analyses in time and frequency domain are not independent but cross check one another using different signal and background averaging and subtraction methods.  Because there is a higher number of measurements passing selection criteria in the time domain analysis than in the frequency domain analysis (see Table 2), we quote this result as our best estimate of the attenuation coefficient:

\begin{equation}
\left< \alpha \right> = 3.20 \pm 0.57~{\rm km^{-1} }.
\end{equation}

\noindent which expressed as an attenuation length is:

\begin{equation}
\left< \lambda \right> = 312^{+68}_{-47}~{\rm m}.
\end{equation}

\noindent The results of all analyses are consistent with an attenuation length of $\sim$300 m $\pm$ 20\,\%.

Up to 30 kHz no strong frequency dependence of $\alpha$ has been found. There are also no indications of depth dependence of the attenuation up to 500 m depth.

\begin{table*}
\resizebox{0.95\textwidth}{!}{
\begin{minipage}[h]{0.95\linewidth}
\caption{\label{systematicsTable}Systematic effects present in individual analyses.}
\begin{center}
\begin{tabular}{ p{.6\textwidth}  p{.1\textwidth}  p{.1\textwidth} p{.1\textwidth}  p{.1\textwidth}}
Systematic Effect 				& pinger	& inter-string single level& inter-string ratio & transients\\ 
\hline  
\hline
Channel-to-channel sensitivity variation	& no 	& yes 	& no & minimal \\     
Azimuthal sensitivity variation		& minimal  	& yes 	& yes & minimal \\
Polar sensitivity variation 		& minimal 	& no 	& yes & minimal \\
\hline
Channel-to-channel transmittivity variation	& no 	& no 	& no & no \\     
Azimuthal transmittivity variation	& no 	& yes 	& yes & no \\
Polar transmittivity variation 		& no 	& no	& yes & no \\
\hline
\end{tabular}
\end{center}
\end{minipage}
}
\end{table*}

Our measured value for the attenuation coefficient is an order of magnitude larger than expected.  In~\cite{PR06}, it was estimated that South Pole ice grains are sufficiently small that Rayleigh scattering is negligible and attenuation is dominated by absorption due to proton reorientation.  However, new data~\cite{ME09} from the SPRESSO site near the South Pole indicate that the ice grains are larger than previously estimated.  Because Rayleigh scattering increases with the cubic power of the grain length, it could be that the attenuation we have measured is dominated by scattering.  Another possible mechanism of absorption, not previously considered, is internal friction at linear crystallographic dislocations.

The weak frequency dependence observed in our measurements below 30 kHz disfavors the Rayleigh scattering hypothesis.  New pinger measurements have been taken in the 2009-2010 South Pole summer in order to clarify the situation. A modified pinger, emitting a sequence of acoustic pulses at three different frequencies up to 60 kHz, was deployed up to 1000 m depth in three additional bore holes. Analysis of these data is currently underway, and should allow for a more conclusive study of both the frequency and depth dependence of acoustic attenuation in South Pole ice.   

Given some of the inherent advantages of the acoustic technique relative to the radio technique (such better shielding from anthropogenic surface backgrounds), this study was undertaken primarily to determine whether the acoustic method could be a basic ingredient of a 100 km$^3$ hybrid detector for ultra-high energy neutrino detection at the South Pole. The design chosen in \cite{BE05} with horizontal string distances of order 1~km will not reach the necessary sensitivity, given the 300~m attenuation length reported here. New geometries are under study to assess whether detectors with closer string spacing but larger area at smaller depth could be an alternative solution.  Measurements are also in preparation to measure the absolute noise level of the Gaussian noise, which is the most important remaining ingredient to determine the feasibility of the acoustic method for detecting neutrinos in South Pole ice.


\section{Acknowledgments}
We acknowledge the support from the following agencies: U.S. National Science Foundation-Office of Polar Program, U.S. National Science Foundation-Physics Division, University of Wisconsin Alumni Research Foundation, U.S. Department of Energy, and National Energy Research Scientific Computing Center, the Louisiana Optical Network Initiative (LONI) grid computing resources; Swedish Research Council, Swedish Polar Research Secretariat, Swedish National Infrastructure for Computing (SNIC), and Knut and Alice Wallenberg Foundation, Sweden; German Ministry for Education and Research (BMBF), Deutsche Forschungsgemeinschaft (DFG), Research Department of Plasmas with Complex Interactions (Bochum), Germany; Fund for Scientific Research (FNRS-FWO), FWO Odysseus programme, Flanders Institute to encourage scientific and technological research in industry (IWT), Belgian Federal Science Policy Office (Belspo); Marsden Fund, New Zealand; Japan Society for Promotion of Science (JSPS); the Swiss National Science Foundation (SNSF), Switzerland; A. Kappes and A. Gro{\ss} acknowledge support by the EU Marie Curie OIF Program; J. P. Rodrigues acknowledge support by the Capes Foundation, Ministry of Education of Brazil.



\bibliography{acoustic_attenuation}
\bibliographystyle{elsart-num}

\end{document}